\documentclass[12pt]{article}
\usepackage{amsmath}
\usepackage{amsfonts}
\newcommand{\R}{\mathbb{R}}

\newcommand{\p}{\partial}

\begin{document}
\begin{titlepage}
\begin{center}
{\large \bf On Certain Quantization Aspects\\of
\\(Generalized) Toda Systems}
\vskip 4em
{\large M. L\'egar\'e}\\
\vskip 1em
Department of Mathematical and Statistical Sciences\\
University of Alberta\\
Edmonton, Alberta, Canada, T6G 2G1\\
mlegare@math.ualberta.ca\\
ph : (780)-492-0567, fax : (780)-492-6826
\end{center}
\vskip 2em

\begin{abstract}
Ordinary and $gl(n,\R)$ generalized Toda systems as well as a related hierarchy are probed with respect to certain quantization characteristics. ``Quantum" canonical and Poisson transformations are used to study quantizations of transformed Toda systems. With a Lax pair setting, a hierarchy of related systems are shown and their quantizations discussed. Finally, comments are added about quantum aspects of $gl(n,\R)$ generalized Toda systems with the approaches of deformation quantization or quantum groups in mind.

\noindent\sloppy{{\bf PACS} : 02.30.Ik,03.65.-w}

\noindent\sloppy{{\bf MSC} : 37N20,81S30}

\noindent\sloppy{{\bf Keywords} : Toda Systems, Quantization, Nonlinear Differential Equations}

\end{abstract}
\end{titlepage}

\pagebreak

\noindent
{\bf Introduction}

In this short communication, certain quantization aspects and properties of some (ordinary, i.e. classical finite) open Toda as well as $gl(n,\R)$ generalized  Toda models are considered. (Ordinary) open Toda lattice systems have been studied with different approaches to quantization in various articles already (see for instance \cite{GW,OP2,STS} and references therein) to derive for example Schr\"odinger equations and eigenfunctions. In this article, two main directions have been followed. First, the use of ``quantum" canonical transformations \cite{A1,A2} in toy model generalizations of the $A_1$ Toda systems and in the $A_2$ Toda model has been made. Most results on equivalent quantizations that have been obtained here would be extendable to $A_{N-1}, N >3$ Toda systems as well. They include the introduction and some aspects of ``quantum" Poisson transformations, which can be seen as extensions of canonical ones. 

Then a related $A_2$ Toda hierarchy is defined and its quantization (along certain preliminary steps of the canonical, geometric, and deformation approaches, see \cite{AE}) has been probed along the lines developed in the previous section (that is, using quantum canonical and Poisson transformations). Here too, such a hierarchy could be extended to larger $A_{N-1}$ cases, but further steps in its quantization would have to be investigated further and seem more difficult to carry out. 

In view of these results, discussions and comments have been presented for the quantization of $gl(n,\R)$ generalized Toda systems. After a Hamiltonian formulation of $gl(n,\R)$ generalized Toda systems, discussions of two possible quantization approaches are carried out as a second direction : deformation quantization \cite{AE,BGGS,St} and quantum groups \cite{BGGS,Ca,Takh,CP}. Some details are provided on the deformation quantization approach, which would appear more realizable. The construction of a $\ast$ - product is described, given the linear Poisson brackets associated with the Hamiltonian structure of such $gl(n,\R)$ systems. The use of quantum groups is also mentioned and some comments to produce an explicit $\ast$ - product in this context are shortly recalled.  

\medskip
\noindent
{\bf Generalized $\mathbf A_1$ Toda (Toy) Models}

The $A_1$ (also called $N=2$ open or nonperiodic) Toda (\cite{Ho,LS,OP1,T1}) system has the following Hamiltonian ($H$) in the center of mass system :
\begin{equation}\label{H-cm}
H(P,Q) = \dfrac{P^2 + e^{2Q}}{2},
\end{equation}
with $\omega_{(P,Q)} = dP \wedge dQ$.

The dynamical equation for $Q$ can also be written in terms of a variable $\tilde q$, using the transformation : $\tilde q = e^Q$, with :
\begin{equation}\label{tilde-H-cm}
H(\tilde p,\tilde q) = \dfrac{\tilde p^2 + \tilde q^2}{2}
\end{equation}
and the symplectic form : $\omega_{(\tilde p,\tilde q)} = \dfrac{d\tilde p \wedge d\tilde q}{\tilde q}, \tilde q \neq 0$.

 A generating function of the type \cite{GPS} $F_2(q,P,t) = P \ln(q)$, would canonically transform the Hamiltonian (\ref{H-cm}) into the following expression :
\begin{equation}\label{new-H-cm}
H(p,q) = \dfrac{p^2q^2 +q^2}{2}
\end{equation}
where $Q=\ln(q)$ and $P=pq$. The same dynamical equations are derived again with $\omega_{(p,q)} = dp\wedge dq$ (see \cite{Le}).

A quantization of the canonical type \cite{AE,Ma}, called ``standard" in what follows, of system (\ref{H-cm}) involves : $[\hat P, \hat Q] = -i, \hat P = -i\dfrac{d}{d Q}, \hat Q = Q$, and brings the following stationary Schr\"odinger equation (setting $\hbar = 1$ in this article) :
\begin{equation}\label{schrodinger-1}
\dfrac{1}{2}\left [-\dfrac{d^2}{d Q^2} + e^{2Q} \right ] \Psi(Q) = E \Psi(Q),
\end{equation}
with energy $E$, for which solutions can for example be found in terms of Bessel functions, given finite conditions at $\pm \infty$ on $\Psi$ (see \cite{OP2}).

However, a quantization through the Hamiltonian description (\ref{tilde-H-cm}) using $\omega_{(\tilde p,\tilde q)}$ along with the following quantization conditions and realization for the operators $\hat{\tilde p}, \hat{\tilde q}$ : $[\hat{\tilde p},\hat{\tilde q}] = -i\hat{\tilde q}, \hat{\tilde p} = -i\tilde q \dfrac{d}{d \tilde q}, \hat{\tilde q} = \tilde q$, brings an equivalent stationary Schr\"odinger equation via the transformation : $\tilde q = e^Q$, on equation (\ref{schrodinger-1}). The conditions imposed on $\tilde \psi(\tilde q)$ would have to agree with those imposed on $\Psi(Q)$. (See for instance \cite{MTdC} for different symplectic settings in certain problems.) 

The third Hamiltonian description (\ref{new-H-cm}) leaves a Hamiltonian in terms of the variables $p$ and $q$. Since it has been obtained via a canonical transformation, it would lead to standard commutation relations and representation for $\hat p$ and $\hat q$ ($[\hat p,\hat q] = -i$, $\hat p = -i \dfrac{d}{d q}, \hat q = q$). An ordering could be necessary and could lead to different quantizations. 

From the Hamiltonian operator : $\hat H = \dfrac{1}{2}(\hat q\hat p\hat q\hat p + {\hat q}^2)$, defined through the (quantum canonical (see \cite{A1,A2})) transformation : $C_r\hat P C_r^{-1} = \hat q \hat p$, where one notes a right-$\hat p$ position, a right-$\hat p$ ordering for $\hat H$ leads to : $\hat H = \dfrac{1}{2}({\hat q}^2{\hat p}^2 - i\hat q\hat p + {\hat q}^2)$, and the corresponding stationary Schr\"odinger equation :
\begin{equation}\label{schrodinger-2}
-q^2 d_q^2\psi_r - qd_q\psi_r + q^2\psi_r =  2 E_r \psi_r
\end{equation}
where $\psi_r(q)$ stands for the wave function and $E_r$, for the energy. The equation (\ref{schrodinger-2}) is equivalent to the stationary Schr\"odinger equation (\ref{schrodinger-1}) with the transformation $q=e^Q$. As expected, a left-$\hat p$ ordered  $\hat H$ above brings the same Schr\"odinger equation (\ref{schrodinger-1}) using $q=e^Q$ again. It is mentioned that either a right-$\hat p$ or left-$\hat p$ ordering on the Hamiltonian operator : $\hat H = \dfrac{1}{2}(\hat p\hat q\hat p\hat q + {\hat q}^2)$, the latter being derived with a left-$\hat p$ ordered (quantum canonical) transformation : $C_l\hat P C_l^{-1} = \hat p \hat q$, leads to a different stationary Schr\"odinger equation ($q = e^Q$). Also, a Weyl (or symmetric) ordering (\cite{AE,deG,deGS}) will be associated with same boundary conditions with the following Hamiltonian :
\begin{equation}
\hat H = \dfrac{1}{8}\left [{\hat q}^2{\hat p}^2 + 2\hat p{\hat q}^2\hat p + {\hat p}^2{\hat q}^2\right ] + \dfrac{1}{2} {\hat q}^2
\end{equation}
The corresponding stationary Schr\"odinger equation can be seen as the (average) sum of the stationary Schr\"odinger equations derived from the $C_r\hat P C_r^{-1}$ and $C_l\hat P C_l^{-1}$ (quantum canonical) transformations. The previous $C_r$ and $C_l$ transformations led to transformations labelled as quantum canonical transformations (\cite{A1,A2}) and preserve the commutation relations : $[\hat P,\hat Q] = -i$, $[\hat p,\hat q] =-i$, other fundamental brackets vanishing. It is known that no rule can in general be drawn for a correspondence between classical and quantum canonical transformations (\cite{A1,A2}).

Differently, the equation of motion, or dynamical equation, for $q(t)$ in the above system with $H$ equation (\ref{new-H-cm}) can be obtained via the Lax pair \cite{Fl} :
\begin{equation}\label{general-lax}
L = \left [\begin{matrix} p(t) & q(t)\\ q(t) & -p(t) \end{matrix} \right ], \quad M = \dfrac{1}{2}\left [\begin{matrix} 0 & -q(t) \\ q(t) & 0 \end{matrix} \right ],
\end{equation}
with the Lax equation :
\begin{equation}
\dfrac{dL}{dt} = [L,M]
\end{equation}

A toy generalization of such systems can be written, using the previous Lax matrix with a more general auxiliary $M$ to form the following Lax pair \cite{Le} :
\begin{equation}
L = \left [\begin{matrix} p & q\\ q & -p \end{matrix} \right ], \quad M = \dfrac{1}{2}\left [\begin{matrix} 0 & -q^n \\ q^n & 0 \end{matrix} \right ],
\end{equation}
with the Lax equation $\dfrac{dL}{dt} = [L,M]$, where $n$ is any integer.

Two Hamiltonians descriptions of the corresponding dynamical equations for the Lax systems (\ref{general-lax}) have the form :
\begin{alignat}{3}\label{H-forms}
&(a)\quad & H=&\dfrac{1}{2}(p^2 q^{2n} +q^2),\quad &\text{with} \quad \omega &= dp\wedge dq\nonumber \\
&(b)\quad & \tilde H=&\dfrac{1}{2}({\tilde p}^2 +{\tilde q}^2), &\text{with} \quad \tilde\omega &= \dfrac{d\tilde p\wedge d\tilde q}{{\tilde q}^n} 
\end{alignat}

One notes that the above Lax equations can be obtained from the oscillator Lax set ($n=0$) by introducing a factor $q^n$ between the Poisson brackets, since $\{H,L\}_{\omega} \propto [L,M]$.
A quantization could proceed with either (nonvanishing fundamental commutation relations shown only)  (a) standard $[\hat p,\hat q] = -i, \hat p = -i \dfrac{d}{d q}, \hat q = q$, or (b) $[\hat{\tilde p},\hat{\tilde q}] = -i {(\hat{\tilde q})}^n, \hat{\tilde p} = -i{\tilde q}^n \dfrac{d}{d q}, \hat{\tilde q} = \tilde q$, leading to the following stationary Schr\"odinger equation (in either $q$ or $\tilde q$ variable) :
\begin{equation}\label{schrodinger-qn}
\dfrac{1}{2}\left [-i q^n \dfrac{d}{d q}\left (-i q^n \dfrac{d}{d q} \right ) + q^2 \right ] \psi(q) = \epsilon \psi(q)
\end{equation}
where for the Hamiltonian formulation (a) above, the (quantum) Hamiltonian is given by : $\hat H = \dfrac{1}{2}({\hat q}^n \hat p {\hat q}^n \hat p + {\hat q}^2)$. In analogy with quantum canonical transformations, an interpretation of the relation between the quantum relations associated to the phase spaces in the formulations (a) and (b) in equation (\ref{H-forms}), which classically is realized as a Poisson map, could be called quantum Poisson map (or transformation). Similarly to quantum canonical transformations, such transformations can be defined independently from a Hilbert space structure and Hamiltonian (\cite{A1,A2}), but a physical equivalence of the quantum formulations obtained, as for canonical transformations, would have to be explored.

One notes that the Schr\"odinger equation (\ref{schrodinger-qn}) can take a simpler linear form via a change of variable for $n \neq 1$ :
$u = \dfrac{1}{1-n}q^{(n-1)}$,
which is equivalent to a quantum canonical transformation:
\begin{equation}
\hat U = {\hat q}^n\hat p, \quad \hat u = \dfrac{1}{1-n} {\hat q}^{1-n}
\end{equation}
applied on the Hamiltonian ($H$) of the formulation (a) in equation (\ref{H-forms}). The Schr\"odinger equation then takes the form :
\begin{equation}
\dfrac{1}{2}\left [-\dfrac{d^2}{d u^2} + \dfrac{1}{[(1-n) u]^{(2/(n-1))}} \right ] \psi(u) = \tilde\epsilon\psi(u)
\end{equation}

It is recalled that the $n=1$ case has been discussed previously. As for $n = -1, 0, 2,3$, they respectively bring potentials of the linear, oscillator, Calogero and electrostatic types. (One could add that the above equation can be related to a particular form of the Riccati equation \cite{Ka}, and that solutions can be given using known series methods for many values of $n$.) Suitable boundary conditions would normally have to be imposed on the solutions found as well. It would be of interest to study quantum canonical and Poisson transformations for additional constants of motion of the $A_n$ Toda models, known to be superintegrable \cite{ADS}.
   
\medskip
\noindent
{\bf $\mathbf A_2$ Toda Systems and a Quantum Canonical Transformation}

The $A_2$ (also called $N=3$ open or nonperiodic) Toda system has a following Hamiltonian \cite{OP2} :
\begin{equation}\label{H-a2}
H(p_i,x_i) = \dfrac{1}{2}(p_1^2+p_2^2+p_3^2) + e^{(x_1-x_2)} + e^{(x_2-x_3)}
\end{equation}
with the symplectic form : $\omega_{(p_i,x_i)} = \sum\limits_{i=1}^3 dp_i\wedge dx_i$. Because of the center of mass motion, one can introduce a projection map (change of variables) :
\begin{equation}
\xi=x_1-x_2, \quad \eta = x_2-x_3,
\quad p_\xi = \dfrac{1}{3}(2\dot\xi + \dot\eta),\quad p_\eta = \dfrac{1}{3}(\dot\xi +2\dot\eta)
\end{equation}
which brings the Hamiltonian :
\begin{equation}
H(p_\xi,p_\eta,\xi,\eta) = p_\xi^2 -p_\xi p_\eta + p_\eta^2 + e^\xi + e^\eta
\end{equation}
with the transformed symplectic form : $\omega_{(p_\xi,p_\eta,\xi,\eta)} = dp_\xi\wedge  d\xi + dp_\eta\wedge d\eta$. Note that below a dotted variable indicates a derivative of this variable with respect to the time variable.

A quantization step could be provided with the following standard fundamental commu\-tation relations (nonvanishing) and representations : \newline $[\hat p_\xi, \hat\xi] = -i, \hat p_\xi =-i \p_\xi, \hat\xi = \xi, [\hat p_\eta, \hat\eta] = -i, \hat p_\eta =-i \p_\eta, \hat\eta = \eta$.

Solutions to a Schr\"odinger equation in related sets of coordinates and momentum representations have been given in references \cite{OP2,BLOPR}. It is (recalled) mentioned that a transformation to Jacobi coordinates \cite{OP2,BLOPR} provides a description of the center of mass and of the motion with respect to the center of mass. A classical canonical transformation would link to a formulation in the center of mass in terms of generalized coordinates $\xi,\eta$. Quantum mechanically, this change can be implemented using a quantum canonical transformation. (The center of mass Hamiltonian and its quantization (contributing to the eigenfunctions as part of a tensor product) being ignored in what follows.)

An interest is to carry out, analogously to the $A_1$ Toda system, ``quantum Poisson and canonical" transformations to a completely integrable \\(higher dimensional) system possessing two (commuting) conserved quantities.

First, with in mind higher dimensional cases (where the following could be useful), one considers a Lax pair formulation of the system (\ref{H-a2}) \cite{Ho,KV} :
\begin{equation}\label{lax-pair}
L = \left [\begin{matrix} v_0 & c_0 & 0\\ c_0 & v_1 & c_1 \\ 0 & c_1 & v_2 \end{matrix} \right ], \quad A_0 = \dfrac{1}{2}\left [\begin{matrix} 0 & c_0 & 0 \\ -c_0 & 0 & c_1 \\ 0 & -c_1 & 0 \end{matrix} \right ],
\end{equation}
with Lax equation : $\dot L = [A_0,L]$, where :
\begin{equation}
c_0 = e^{(x_1-x_2)/2}, c_1 = e^{(x_2-x_3)/2}, v_0 = -p_1, v_1 = -p_2, v_2 = -p_3
\end{equation}
Then a Hamiltonian : $H = \dfrac{1}{2}\text{tr}(L^2)$, and a second invariant : $I  = \text{tr}(L^3)$ commuting with $H$ (with respect to $\omega_{(p_i,x_i)}$ or $\omega_{(p_\xi,p_\eta,\xi,\eta)}$) are chosen, besides the invariant $\text{tr}(L)$. Explicitly, $I$ has the form :
\begin{align}
I &= -p_1^2-p_2^2-p_3^2 -3e^{(x_1-x_2)}(p_1+p_2) - 3 e^{(x_2-x_3)}(p_2+p_3)\\
I &= 3[p_\eta^2 p_\xi - p_\eta p_\xi^2 - e^\xi p_\eta + e^\eta p_\xi]
\end{align}

Using standard quantization relations and representation for the ($p_i,x_i$) phase space, as well as for the ${(p_\xi,p_\eta,\xi,\eta)}$ phase space, it is seen that no ordering choice is needed to define quantum versions of $H$ and $I$, here denoted $\hat H$ and $\hat I$, respectively. It is also checked that, as expected : $[\hat H,\hat I] = 0$ (see \cite{Gu,Hi}).

Along with the $A_1$ Toda model development in the above section, a canonical transformation is introduced :
\begin{equation}
Q_1 = e^{\xi/2}, Q_2 = e^{\eta/2}, P_1 = 2p_\xi e^{-\xi/2}, P_2 = 2p_\eta e^{-\eta/2}
\end{equation}
which leads to :
\begin{equation}
H(Q_1,Q_2,P_1,P_2) = \dfrac{1}{4}[Q_1^2P_1^2 + Q_2^2P_2^2 -Q_1Q_2P_1P_2] + Q_1^2+ Q_2^2
\end{equation}
where $\omega_{(P_1,P_2,Q_1,Q_2)} = dP_1\wedge dQ_1 + dP_2\wedge dQ_2$.

One can give a standard quantization in terms of the new variables
$[\hat P_1,\hat Q_1] = -i, [\hat P_2,\hat Q_2] = -i, \hat P_1 = -i \p _{Q_1}, \hat P_2 = -i \p_{Q_2}, \hat Q_1 = Q_1, \hat Q_2 = Q_2$, other fundamental brackets vanishing. A corresponding quantum canonical transformation (with right-$\hat P_i$ ordering) : 
\begin{equation}\label{can-transf-a2}
C\hat p_\xi C^{-1} = \dfrac{1}{2} \hat Q_1\hat P_1, C \hat p_\eta C^{-1} = \dfrac{1}{2} \hat Q_2\hat P_2, C^{-1}\hat Q_1 C = e^{{\hat \xi}/2}, C^{-1} \hat Q_2 C = e^{{\hat \eta}/2}
\end{equation}
brings the commuting quantum Hamiltonian ($\hat H$) and the second quantum invariant ($\hat I$) :
\begin{align}
\hat H &= \dfrac{1}{4}\left [(\hat Q_1\hat P_1)^2+ (\hat Q_2\hat P_2)^2 - (\hat Q_1\hat P_1)(\hat Q_2\hat P_2)\right ] + (\hat Q_1)^2 + (\hat Q_2)^2 \\
\hat I &=\dfrac{1}{8}\left [(\hat Q_2 \hat P_2)^2(\hat Q_1\hat P_1) - (\hat Q_2\hat P_2) (\hat Q_1\hat P_1)^2 \right ] -\dfrac{1}{2}\hat Q_1^2\hat Q_2\hat P_2 + \dfrac{1}{2}\hat Q_2^2\hat Q_1 \hat P_1 
\end{align}

As found for a quantization of the $A_1$ Toda system with the right-$\hat p$ ordering, the stationary Schr\"o\-dinger equations formulated in the coordinates $(\xi,\eta)$ and $(Q_1,Q_2)$ are equivalent, via a coordinate transformation. A left-$\hat p$ ordering involves different stationary Schr\"odinger equations.

One can add that a Weyl (symmetric) quantization could also be carried out similarly to the $A_1$ Toda system. Interestingly, there also exists a quantum Poisson transformation :
\begin{equation}
P \hat \pi_1 P^{-1} = \dfrac{1}{2}\hat Q_1 \hat P_1, P \hat \pi_2 P^{-1} = \dfrac{1}{2}\hat Q_2 \hat P_2,  P \hat \lambda_1 P^{-1} = \hat Q_1, P \hat \lambda_2 P^{-1} = \hat Q_2
\end{equation}
with $\omega_{(P_1,P_2,Q_1,Q_2)}$ given above and $\Omega = 2\dfrac{d\pi_1\wedge d\lambda_1}{\lambda_1} + 2\dfrac{d\pi_2\wedge d\lambda_2}{\lambda_2}$, leading to :
\begin{equation}
\hat H(\hat\pi_1,\hat\pi_2,\hat\lambda_1,\hat\lambda_2) = {{\hat \pi}_1}^2 -{\hat \pi}_1{\hat \pi}_2 + {{\hat \pi}_2}^2 + {{\hat \lambda}_1}^2 + {{\hat\lambda}_2}^2
\end{equation}
The second invariant would then be written as :
\begin{equation}
\hat I  = {{\hat \pi}_2}^2{\hat \pi}_1 - {\hat \pi}_2{{\hat \pi}_1}^2 - {{\hat \lambda}_1}^2 {\hat\pi}_2 + {{\hat\lambda}_2}^2 {\hat\pi}_1
\end{equation}

Fundamental quantum relations (non-vanishing commutators only) and a representation could be :
\begin{equation}
[{\hat\pi}_1,{\hat\lambda}_1] = \dfrac{-i}{2} {\hat\lambda}_1, \quad [{\hat\pi}_2,{\hat\lambda}_2] = \dfrac{-i}{2} {\hat\lambda}_2,
\end{equation}
\begin{equation} 
{\hat\pi}_1 = -i\lambda_1\p_{\lambda_1}, {\hat\pi}_2 = -i\lambda_2\p_{\lambda_2}, {\hat\lambda}_1 = \lambda_1, {\hat\lambda}_2 = \lambda_2
\end{equation}
A first comment could be made that a generalization of such system through a symplectic form :
\begin{equation}
\Omega_n = 2\dfrac{d\pi_1\wedge d\lambda_1}{{\lambda_1}^n} + 2\dfrac{d\pi_2\wedge d\lambda_2}{{\lambda_2}^n}
\end{equation}
where $n$ is an integer $\neq 1$, in analogy again with the above-mentioned $A_1$ toy generalizations, brings systems for which Liouville integrability may not be satisfied, since at the classical level : $\{H,I\}_{\Omega_n} \neq 0$.

Secondly, quantum canonical transformations similar to equation (\ref{can-transf-a2}) can be built for $A_{N-1}$ Toda systems with $N > 3$ particles using a set of generalized coordinates of the same kind as those elected for $A_2$ : ($\xi,\eta$), starting for instance with Jacobi coordinates.

Finally, since no full set of rules is known allowing to translate classical canonical transfromations in quantum form (see \cite{A1,A2}), (nonlinear) canonical transformations to action-angle variables for such Toda systems (see \cite{DLNT,Mos,vanM}) would have to be investigated each separately \cite{Gu}.
  
\medskip\noindent
{\bf Related $\mathbf A_2$ Toda Hierarchy}

A known Toda hierarchy is obtained via the classical $r$ - matrix : \newline
$r = \sum\limits_{i>j=1}^3(E_{ij}\otimes E_{ji} - E_{ji}\otimes E_{ij})$, given the matrix $L$ in the Lax pair equation (\ref{lax-pair}), where however $A_0$ is defined as :\newline \centerline{$\text{tr}_{(\text{on first factor in}\, \otimes \,\text{product})}((L^m \otimes 1) r)$} or \newline \centerline{$((L^m)_{(\text{upper triangular})} - (L^m)_{(\text{lower triangular})})$} for any positive integer $m$ (see \cite{Ta} and references therein).

To give an idea of difficulties that could be involved in quantizing, a simple hierarchy is considered in what follows, with the hope of generalizing certain results of the above sections. For instance, one can look instead at the following set of Lax equations :
\begin{equation}
\dot L = [A_0^{2n-1},L], \quad n = 1,2,3,...
\end{equation}
where $L$ and  $A_0$ are forming the Lax pair defined by equation (\ref{lax-pair}). The Lax equations obtained are :
\begin{align}
\dot v_0 =& \dfrac{1}{2^{(2n-2)}} (-1)^{(n+1)} c_0^2 (c_0^2 + c_1^2)^{(n-1)} \\
\dot v_1 =& \dfrac{1}{2^{(2n-2)}} (-1)^n (c_0^2 - c_1^2) (c_0^2 + c_1^2)^{(n-1)} \\
\dot v_2 =& \dfrac{1}{2^{(2n-2)}} (-1)^n c_1^2 (c_0^2 + c_1^2)^{(n-1)} \\
\dot c_o =& \dfrac{1}{2^{(2n-1)}} (-1)^n (v_0 - v_1) c_0 (c_0^2 + c_1^2)^{(n-1)}\\
\dot c_1 =& \dfrac{1}{2^{(2n-1)}} (-1)^n (v_1 - v_2) c_1 (c_0^2 + c_1^2)^{(n-1)}
\end{align}
with $\dot v_0 + \dot v_1 +\dot v_2 = 0$, as expected, and $n =1,2,3, ...$

Due to the invariance of these equations under the transformation : $v_\alpha \rightarrow v_\alpha + c$, with $\alpha = 0,1,2$, one can select : $v_0 + v_1 + v_2  = 0$. Thus, a projection map is given as :
\begin{equation}
c_0 = e^{\xi/2}, c_1 = e^{\eta/2},\quad \text{and}\; v_0 = -p_\xi, v_1 = -p_\eta + p_\xi, v_2 = p_\eta
\end{equation}
It follows that the Lax equations become :
\begin{align}
\dot p_\xi &= \dfrac{1}{2^{(2n-2)}} (-1)^n e^\xi (e^\xi + e^\eta)^{(n-1)}\\
\dot p_\eta &= \dfrac{1}{2^{(2n-2)}} (-1)^n e^\eta (e^\xi + e^\eta)^{(n-1)}\\
\dot \xi &= \dfrac{1}{2^{(2n-2)}} (-1)^{(n+1)}(2p_\xi - p_\eta) (e^\xi + e^\eta)^{(n-1)}\\
\dot \eta &= \dfrac{1}{2^{(2n-2)}} (-1)^{(n+1)}(2p_\eta - p_\xi) (e^\xi + e^\eta)^{(n-1)}
\end{align}
These equations could be derived as Hamilton equations, similarly to the $A_2$ Toda system presented in the previous section :
\begin{equation}
H = \dfrac{1}{2}\text{tr}(L^2) = (p_\xi)^2 - p_\xi p_\eta + (p_\eta)^2 + e^\xi + e^\eta
\end{equation}
equipped with the Poisson brackets, all other fundamental brackets vanishing :
\begin{equation}\label{p-b}
\{p_\xi,\xi\} = \dfrac{(-1)^{n+1}}{2^{(2n-2)}}(e^\xi + e^\eta)^{(n-1)} \quad \{p_\eta,\eta\} = \dfrac{(-1)^{n+1}}{2^{(2n-2)}}(e^\xi + e^\eta)^{(n-1)}
\end{equation}
A corresponding 2-form ($\omega$) in terms of $\xi,\eta, p_\xi, p_\eta$ would not be symplectic, and Darboux coordinates could not be chosen. Nevertheless, a quantization could involve applying the deformation quantization approach to derive a $\ast$ product, or attempting a set of quantum brackets modeled on the Poisson brackets above (equation (\ref{p-b}))(see \cite{AE} and references therein).

It is added that such hierarchies of Lax equations are expected to be derived for $A_{N-1}, N >3$ Toda systems generalizations as well, using the odd powers of the auxiliary matrix $A_0$ .

\medskip\noindent
{\bf Generalized Nonperiodic Toda Systems}

 A $gl(2,\R)$ - Toda system can be generated with the following Lax equations : $\dot L = [P,L]$ (\cite{DLT,KY}, as Cholesky flows) :
\begin{equation}\label{lax-gl2}
L = \left [\begin{matrix} a_{11} & a_{12}\\ a_{21} & a_{22} \end{matrix} \right ], \quad P = \left [\begin{matrix} 0 & a_{12} \\ -a_{21} & 0 \end{matrix} \right ],
\end{equation}
where $L$ belongs to $gl(2,\R)$.

Equivalently, these equations can be derived via a Hamiltonian formulation using as nondegenerate invariant scalar product : $<L,M> = \text{tr}(LM)$, for $L,M \in gl(n,\R)$, and the $R$ - matrix : $R = \pi_+ - \pi_-$, where $\pi_+, \pi_-$ correspond respectively to (strictly) upper and lower triangular projections of $gl(n,\R)$. Using a basis $E_{ij}, i,j = 1,...,n$ for $gl(n,\R)$, where $[E_{ij}]_{kl} = \delta_{ik}\delta_{jl}$, with thus : $L = \sum\limits_{i,j = 1}^n L_{ij}E_{ij}$, a Lie-Poisson bracket can be defined as a linear $r$ - matrix bracket associated to the $R$ - matrix \cite{Su} :
\begin{equation}\label{pb-gl}
2\{L_{ij},L_{kl}\} = <[R(E_{ji}), E_{lk}] + [E_{ji}, R(E_{lk})],L>,
\end{equation}
where $i,j,k,l =  1,2,...,n$.

For the $gl(2,\R)$ (thus $n=2$) generalized nonperiodic (or open) Toda system discussed below, choosing the Hamiltonian : $H = - \text{tr}(L^2) = - (a_{11}^2 + a_{12}^2 + 2 a_{12}a_{21})$, gives rise to the Lax equations of (\ref{lax-gl2}) as Hamilton equations :
\begin{equation}
\dot{a_{ij}} = \{H,a_{ij}\}
\end{equation}
Note that $\text{tr}(L^3)$ is functionally dependent on $\text{tr}(L)$ and $\text{tr}(L^2)$ ($\text{tr}(L^3) = - (\text{tr}(L))^3 + 3 \text{tr}(L^2) \text{tr}(L)$), but $\text{tr}(L)$ could be associated to a conservation of momentum by keeping a phase space description close to the $A_1$ Toda model formulation with definitions : $a_{11} = p_1, a_{22} = p_2, a_{12} = q_1, a_{21} = q_2$. The Hamiltonian ($H$) then takes the form :
\begin{equation}\label{H-gl2}
H =- [(p_1)^2 + (p_2)^2] -2  q_1q_2
\end{equation}
with fundamental Poisson brackets :
\begin{alignat}{3}
\{p_1,q_1\} &= \dfrac{q_1}{2},&\quad \{p_1,q_2\} &= \dfrac{q_2}{2},&\quad \{q_1,q_2\} &= 0 \\
\{p_2,q_1\} &= \dfrac{-q_1}{2},&\quad \{p_2,q_2\} &= \dfrac{-q_2}{2},&\quad \{p_1,p_2\} &= 0 
\end{alignat}

Quantum corresponding elements could be chosen as :
\begin{equation}
\hat p_1 = -i\left [\dfrac{q_1}{2}\dfrac{\p}{\p q_1} + \dfrac{q_2}{2}\dfrac{\p}{\p q_2}\right ],
\hat p_2 = i\left [\dfrac{q_1}{2}\dfrac{\p}{\p q_1} + \dfrac{q_2}{2}\dfrac{\p}{\p q_2} \right ] + C \hat 1, \hat q_1 = q_1, \hat q_2 = q_2
\end{equation}
where $\text{tr}(L) = p_1 + p_2$ is a constant of motion ($C$). A stationary Schr\"odinger equation would follow from the above defined $H$ (\ref{H-gl2}).

The $gl(3,\R)$ generalized nonperiodic Toda equations can also be obtained using the Hamilton equations : $\dot{a_{ij}} = \{H,a_{ij}\}, i,j = 1,2,3$, and the Hamiltonian : $H = - \text{tr}(L^2)$, with the Poisson brackets of equation (\ref{pb-gl}). One then gets expressions, here presented for any $n \in \mathbb{N}$ with $i,j,k,l = 1,...,n$ :
\begin{align}\label{pb-gln}
\{L_{ii},L_{ll}\} &= 0, \nonumber\\
\{L_{ij},L_{ij}\} &= 0,\nonumber\\
\{L_{ij},L_{kl}\} &= -(\delta_{il} L_{kj} - \delta_{jk} L_{il}), i<j, k<l\nonumber\\
\{L_{ij},L_{kl}\} &= 0, i<j, l<k\\
\{L_{ij},L_{kl}\} &= 0, j<i, k<l\nonumber\\
\{L_{ij},L_{kl}\} &= (\delta_{il} L_{kj} - \delta_{jk} L_{il}), j<i, k>l\nonumber\\
\{L_{ii},L_{kl}\} &= -\dfrac{1}{2}(\delta_{il} L_{ki} - \delta_{ik} L_{il}),  k<l\nonumber\\
\{L_{ii},L_{kl}\} &= \dfrac{1}{2}(\delta_{il} L_{ki} - \delta_{ik} L_{il}),  k>l\nonumber
\end{align}
A standard quantization prescription : $\{\cdot,\cdot\} \rightarrow i [\hat{\cdot},\hat{\cdot}]$ ($\hbar =1$) with linear operators instead of classical functions leaves a structure of solvable Lie algebra. One also stresses that the Poisson tensor is linear in the elements of $L$ ($L_{ij}$).

A set of comments could here be made. Firstly, that transformations leading to action-angle variables on symplectic subspaces of the $\{L_{ij}\}$ space can be written under certain conditions (see \cite{DLT}), but that the quantum version appears to be more involved and could not be directly related to the classical expressions, as for previously mentioned canonical transformations \cite{A1,A2}.

Secondly, an infinite - dimensional representation of the $n^2$ - dimensional (solvable) Lie algebra (equation (\ref{pb-gln})) in terms of operators ($\hat{L_{ij}}$) on an appropriate (Hilbert) space could provide a suitable quantization (see \cite{AE} and references therein). 

Thirdly, for $gl(n,\R)$, a $\ast$ - product can be (formally) obtained on the Poisson manifold $\R^{n^2}$ endowed with the linear Poisson bracket given below. The Campbell - Baker - Hausdorff (CBH) formula can then be used \cite{Kath} to derive such a product. The structure constants $c^{(rs)}_{(ij)(kl)}$ could be read from the Poisson brackets of equation (\ref{pb-gln}) :
\begin{equation}
[L_{ij},L_{kl}] = c^{(rs)}_{(ij)(kl)} L_{rs},
\end{equation}
for $i,j,k,l,r,s = 1,...,n$. It is recalled that the Poisson bracket was derived using the $gl(n,\R)$ commutation relations with the $R \;(=\pi_+ - \pi_-$) matrix. According to theorem (2.13) of ref. \cite{Kath} (math.QA/9811174), the $\ast$ - product is entirely determined by products : $(L_1)^N \ast L_2$, for any $L_1$ and $L_2$ in the above Lie algebra, for $N \in \mathbb{N}$. Those brings sums with Bernoulli numbers and adjoint actions of the Lie algebra in the $\ast$ - products \cite{Kath}. Presenting such sum does not appear to be much of interest for now (a fortiori in a communication which is meant to be short).

Finally, one can mention that instead of the full $gl(n,\R)$, a system involving only $sl(n,\R)$ could be considered, since an element $ L = L_0 + l \mathbf{1}_n$ of $gl(n,\R)$ would lead to the Lax equations : $\dot{L_0} = [M,L_0]$ and $\dot l = 0$, with an unchanged $R$ - matrix : $R = \pi_+(L_0) - \pi_-(L_0)$.

As for the use of quantum groups, it can be added that quantum $R$ - matrices are expected from the knowledge of a $r$ - matrix and that a relation to a $\ast$ - product might be possible via explicit expressions for twisting elements \cite{Ca,Takh}, which infinitesimally would bring back (as $\hbar \rightarrow 0$) the $r$ - matrix. Here, $r$ is solution to the modified Yang-Baxter equations, $r$ is then also said to be quasi-triangular. A Lie-Poisson structure can be introduced on $gl(n,\R)$ (seen as an Abelian Lie group under addition) via the Lie-Poisson bracket :
\begin{equation}
\{f_1,f_2\}_R = <[R(df_1(\xi)),df_2(\xi)] + [df_1(\xi),R(df_2(\xi))],\xi>
\end{equation}
where $\xi$ belongs to $gl^{\ast}(n,\R)$ and $f_1,f_2 \in C^{\infty}(gl^{\ast}(n,\R))$ \cite{CP}.

It is known that a quantization of a Lie bialgebra defined with the matrix $r$ exists \cite{CP}. A deformed associated co-product using a twisting element would be of interest in deriving an associated $\ast$ - product \cite{CP,Bl}.

It is added that in particular for $sl(n)$, quantum $R$ - matrices can be derived for \cite{ESS} $sl(n) \wedge sl(n)$ valued quasi-triangular $r$ - matrices.

It would be interesting to compare $\ast$ - products obtained, but this is not an objective in the present article.

\medskip
\noindent
{\bf Conclusion :}

Properties of the quantization of open (non-periodic) Toda models have been discussed in this short communication. Quantum realizations, using Schr\"odinger equations, of ordinary (classical finite) $A_1$ and $A_2$ Toda systems have been shown to be related via quantum ``canonical and Poisson" transformations, with similar relations holding for open Toda systems with higher number of degrees of freedom, conjectured to follow. A formulation of a hierarchy of related systems has also been probed with respect to similar constructions, but obstructions arose for these particular systems, suggesting the use of different approaches for their quantizations. Finally, $gl(n,\R)$ generalized Toda systems were discussed from the viewpoint of deformation quantization and quantum groups. A $\ast$ - product has been mentioned in the deformation quantization approach and difficulties with respect to the quantum group formulation mentioned. 

As a future development, it would be for example of interest to consider a second step in the deformation quantization approach by associating (Hilbert space) operators to realize a $\ast$ - product of the $gl(n,\R)$ generalized Toda systems. Properties and aspects of ``quantum Poisson" transformations could also form the object of further studies for different systems, together with investigations of the equivalence of quantum systems under such transformations.

\end{document}